\documentclass[aps,prl,reprint,twocolumn,groupedaddress,superscriptaddress,amsmath,amssymb,amsfonts,floats,floatfix,noraggedbottom,balancelastpage,dvips,showpacs,10pt]{revtex4-1}

\usepackage{epsfig} 
\usepackage{psfrag}
\usepackage{braket}
\usepackage{color}
\usepackage[usenames,dvipsnames]{xcolor}
\usepackage{textcomp}
\usepackage{graphicx}
\usepackage{latexsym}
\usepackage{hyperref}
\hypersetup{dvips,
  pdfauthor={Qingyou Meng, Christopher N. Varney, Hans Fangohr, Egor Babaev},
  pdftitle={ Honeycomb, square, and kagome vortex lattices in superconducting systems with multi-scale inter-vortex interactions},
  letterpaper,
  colorlinks=true,
  linkcolor=blue,
  citecolor=blue,
}

\usepackage{breakurl}

\begin{document}

\title{Honeycomb,  square, and kagom\'e vortex lattices in superconducting systems with multi-scale inter-vortex interactions}  

\author{Qingyou Meng}
\affiliation{Department of Physics, University of Massachusetts, Amherst,
  Massachusetts 01003, USA}
\author{Christopher N. Varney}
\affiliation{Department of Physics, University of West Florida,
  Pensacola, FL 32514, USA}
\author{Hans Fangohr}
\affiliation{Engineering and the Environment, University of
  Southampton SO17 1BJ, UK}
\author{Egor Babaev}
\affiliation{Department of Theoretical Physics, The Royal Institute of
  Technology, SE-10691 Stockholm, Sweden}
\affiliation{Department of Physics, University of Massachusetts, Amherst,
  Massachusetts 01003, USA}
  
\begin{abstract}
  The recent proposal of Romero-Isart {\em et
    al.}~\cite{romero-isart_superconducting_2013} to utilize the
  vortex lattice phases of superconducting materials to prepare a
  lattice for ultra-cold atoms-based quantum emulators, raises the
  need to create and control vortex lattices of different symmetries.
  Here we propose a mechanism by which honeycomb, hexagonal, square,
  and kagom\'e vortex lattices could be created in superconducting
  systems with multi-scale inter-vortex interaction.  Multiple scales
  of the inter-vortex interaction can be created and controlled in
  layered systems made of different superconducting material or with
  differing interlayer spacing.

\end{abstract}


\maketitle

To circumvent the limitations on classical computation, a growing
effort to manipulate and control the behavior of ultracold atomic
gases has led to these systems being used as quantum simulators for a
host of phenomena in condensed matter
physics~\cite{buluta_quantum_2009,cirac_goals_2012}.  A focus of
quantum simulator investigations has been on building Hubbard models
by loading a gas of neutral atoms into optical lattices and tuning the
interaction between the atoms~\cite{bloch_many-body_2008,
  bloch_quantum_2012}.  At present, great strides have been made in
cooling protocols~\cite{weld_spin_2009,weld_thermometry_2010,
  mathy_enlarging_2012}. But the main question, to assess in such
experiments whether the Hubbard model can explain high-$T_c$
superconductivity, remains unanswered.

In order to address this question, better cooling schemes which reduce
the entropy of the quantum simulator are
necessary~\cite{bloch_quantum_2012}. Very recently, Romero-Isart {\em
  et al.}~\cite{romero-isart_superconducting_2013} proposed placing
ultracold atoms in a lattice potential generated by magnetic field of
superconducting vortices in type-2 superconductors and trapping the
atoms near the surface. This new approach aims to decrease the
inter-lattice site distance, making the required regimes
experimentally feasible~\cite{gullans_nanoplasmonic_2012,
  romero-isart_superconducting_2013}. This possibility of a crucially
important application raises the need to create and control vortex
lattices of different symmetries. Although in some exotic cases a
square vortex lattice has been
observed~\cite{aegerter_evidence_1998,riseman_observation_1998}, the
overwhelming majority of vortex lattices in superconductors have
hexagonal symmetry. In order to create a vortex lattice of various
symmetries for quantum emulators, Romero-Isart {\em et
  al.}~\cite{romero-isart_superconducting_2013} proposed pinning the
vortices in arrays of etched
holes/anti-dots~\cite{moshchalkov_nanoscience_2010}. While such vortex
systems have been extensively investigated in superconductivity both
theoretically and experimentally for various pinning array
geometries~\cite{baert_composite_1995,moshchalkov_magnetization_1996,
  rosseel_depinning_1996,morgan_asymmetric_1998,grigorenko_direct_2001,
  grigorenko_symmetry_2003,berdiyorov_vortex_2006,reichhardt_vortex_2007,
  cao_temperature_2009,latimer_vortex_2012}, Romero {\em et
  al.}~\cite{romero-isart_superconducting_2013} note that the
anticipated challenges to implementing the approach are high
requirements for perfection of the vortex lattice and possible
variations and field inhomogeneities in the anti-dot arrays.  In fact,
the interest in self-assembly of kagom\'e and honeycomb structures
goes beyond the recent interest in vortex matter and is intensively
studied in soft condensed matter systems~\cite{chen_directed_2011,
  romano_colloidal_2011,mao_entropy_2013,cates_patchy_2013}.


Here we propose an alternative approach involving multi-component
superconducting systems. Recently there has been interest in
superconductivity with several scales of repulsive and attractive
interaction. In two-band superconductors it is possible to have a
vortex system where the short-range interactions are repulsive while
the long-range interactions are attractive in regimes where one
coherence length is shorter than the magnetic field penetration length
while the second coherence length is larger,
i.e. $\xi_1<\lambda<\xi_2$~\cite{babaev_semi-meissner_2005,
  silaev_microscopic_2011,silaev_microscopic_2012,
  garaud_vortex_2012}. The regime which was recently termed type-1.5
superconductivity in experimental works on MgB$_2$
\cite{moshchalkov_type-1.5_2009, gutierrez_scanning_2012,
  dao_giant_2011} and Sr$_2$RuO$_4$~\cite{hicks_limits_2010,
  ray_muon-spin_2014}. The non-monotonic inter-vortex interaction is
also possible in electromagnetically or proximity-effect coupled
bilayers~\cite{babaev_semi-meissner_2005}.
  

In the two-band superconductor the long-range inter-vortex interaction
energy is given by~\cite{babaev_semi-meissner_2005,
  carlstrom_type-1.5_2011,silaev_microscopic_2011}
\begin{align}
  E_\mathrm{int}=C_B^2K_0\left( \frac{r}{\lambda}\right) - C_1^2{2\pi} K_0
  \left(\frac{r}{\xi_1}\right) - C_2^2K_0
  \left(\frac{r}{\xi_2}\right).
\end{align}
The first term describes inter-vortex repulsion which comes from
magnetic and current-current interaction. The second and third terms
describes attractive interactions from cores overlaps. The two
contributions are due to to coherence lengths.

In Ref.~\onlinecite{varney_hierarchical_2013} it was proposed that in
layered systems multiple repulsive length scales are possible when
different layers have different $\lambda_i$.  For a straight and rigid
vortex line, the long-range interaction is then
\begin{align}
  E_\mathrm{int} = \sum_i C_B{}_i^2K_0\left( \frac{r}{\lambda_i}\right) -
  \sum_jC_j^2{2\pi}K_0\left(\frac{r}{\xi_j}\right).
\label{eq:inn}
\end{align}
Such a system can have various cluster phases due to multi-scale
repulsive interactions~\cite{varney_hierarchical_2013}. Subsequently
some of the phases obtained in simulations where the vortices are
treated as a point-particle~\cite{varney_hierarchical_2013} were also
obtained in simulations of a layered Ginzburg-Landau
model~\cite{komendova_soft_2013}.

\begin{figure}[bt]
  \centering
  \includegraphics[height=0.8\columnwidth]{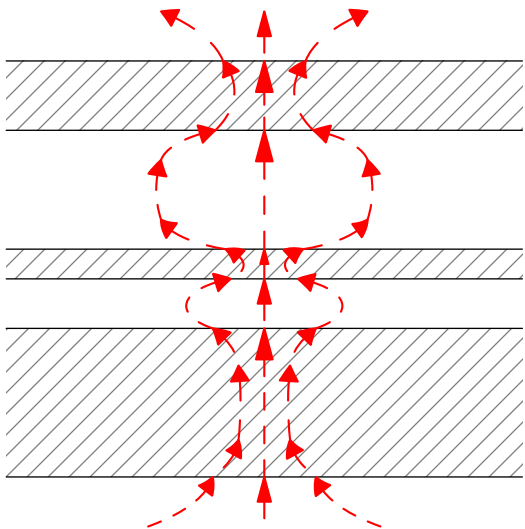}
  \caption{Schematic picture of the magnetic field lines of a vortex
    in a layered superconductor. Shaded (white) areas are
    superconductor (insulator) layers with different thickness. The
    flux spreads in the non-superconducting regions.
    \label{fig:layers} 
  }
\end{figure}
Here we point out that layered systems proposed in
Ref.~\onlinecite{varney_hierarchical_2013}, i.e. structures made of a
combination of type-1 and type-2 superconductors with variable
interlayer distances (see Fig.~\ref{fig:layers}), could be used to
create vortex lattice of different symmetries. In what follows, we
utilize Langevin dynamics to study various states of vortex matter in
superconductors~\cite{fangohr_efficient_2000,fangohr_vortex_2001,
  xu_peak_2008,drocco_static_2013}. Often in systems with multiple
repulsive length scales various phases are quite robust with respect
to potential changes as long as the potential preserves the distinct
repulsive length scales~\cite{malescio_stripe_2004,
  glaser_soft_2007}. Thus we use a phenomenological pairwise potential
with multiple length scales which has characteristic features of the
analytically known asymptotic form Eq.~\eqref{eq:inn} as well as
included effect of demagnetization field in the form of analytically
known long-range power-law repulsive inter-vortex
force~\cite{pearl_current_1964}. We demonstrate that layered systems
where such a potential can be realized can be used to generate the
four two-dimensional lattices: hexagonal, honeycomb, square, and
kagom\'e.

\begin{figure}[bt]
  \centering
  \includegraphics[height=\columnwidth,angle=-90]{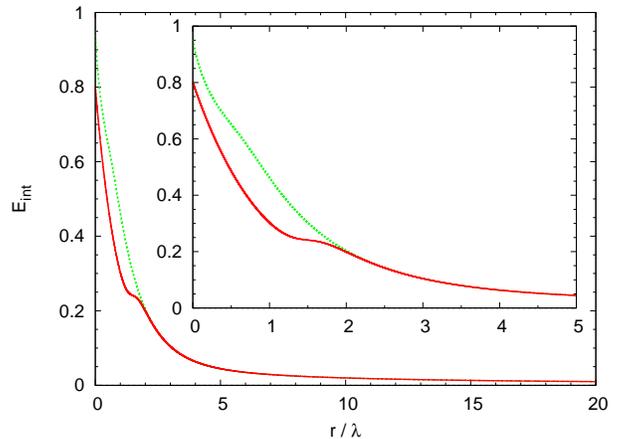}
  \caption{Phenomenological potential that describes the multi-scale
    inter-vortex interaction for straight rigid vortex lines in
    layered system with different layer's parameters. The solid red
    curve gives rise to a honeycomb lattice at density~\cite{noterho}
    $\rho = 1.50$, a hexagonal lattice at $\rho = 2.25$, and a square
    lattice at $\rho = 2.50$, while for the dashed green line a
    kagom\'e lattice is the ground state at a density of $\rho =
    2.50$.~\cite{note1}.
    \label{fig:potentials} 
  }
\end{figure}

In Fig.~\ref{fig:potentials}, we illustrate two potentials that arise
from a phenomenological form
\begin{align}
  E_\mathrm{int} &= c_1 e^{-r/\lambda} - c_2 e^{-r/\xi} + c_3 \frac{\lambda
    \lbrace \tanh[\alpha(r-\beta)]+1\rbrace}{r+\delta}
\end{align}
that captures the essential multi-scale features of the inter-vortex
forces in a layered superconducting
structure~\cite{varney_hierarchical_2013,note1}, when the interaction
can be approximated by pairwise forces between straight vortex
lines. The model features a short-range exponential repulsion,
intermediate-ranged exponential attraction, and a long-range power-law
repulsive behavior. The interplay between these different interactions
results in a rich phase diagram which go beyond the scope of this
paper; we defer a full discussion of its properties for future
work~\cite{meng_longcoldatom}.

\begin{figure}[tb]
  \centering
  \includegraphics[width=0.5\textwidth]{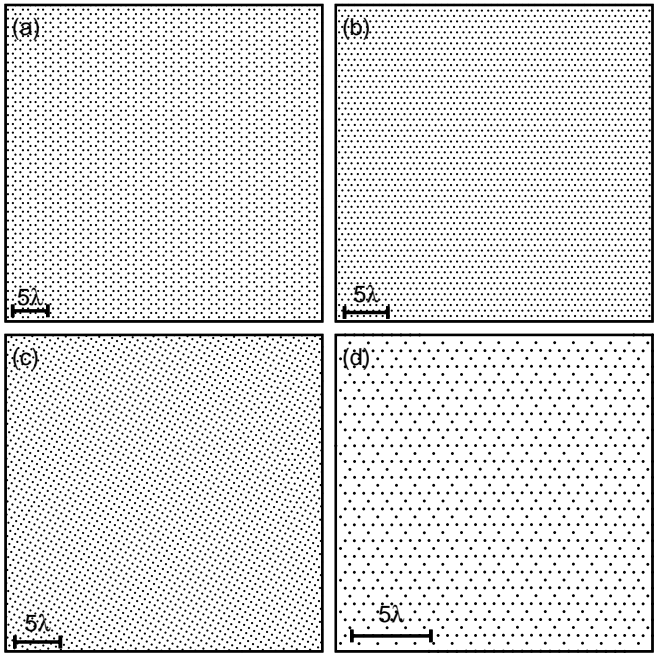}
  \caption{The final vortex configuration at the zero temperature for
    (a) $N_v = 3024$ and $\rho = 1.50$ (honeycomb lattice), (b) $N_v =
    2958$ and $\rho = 2.25$ (hexagonal lattice), (c) $N_v = 2958$ and
    $\rho = 2.50$ (square lattice), and (d) $N_v = 1020$ and $\rho =
    2.50$ (kagom\'e lattice). Panels (a)-(c) correspond to the solid
    red curve of Fig.~\ref{fig:potentials}, while panel (d)
    corresponds to the dashed green curve.
    \label{fig:snapshots} 
  }
\end{figure}

In Fig.~\ref{fig:snapshots}, we illustrate some of the ground state
vortex phases of the potentials shown in
Fig.~\ref{fig:potentials}. The phases were obtained using Langevin
dynamics~\cite{fangohr_vortex_2001} simulations of $N_v \approx 1000$
to $N_v \approx 3000$ vortices where the temperature was slowly
reduced to $T = 0$ (see Refs.~\onlinecite{varney_hierarchical_2013}
and \onlinecite{meng_longcoldatom} for additional details). For the
solid red line of Fig.~\ref{fig:potentials}, we obtain honeycomb,
hexagonal, and square lattices at densities~\cite{noterho} $\rho =
1.50$, $2.25$, and $2.50$, respectively. For the dashed green curve,
we obtain a perfect kagom\'e lattice for $\rho = 2.50$. For the
honeycomb, hexagonal and square lattice results, we find little to no
defects for the largest system sizes studied. For the kagom\'e lattice
results, we achieve a defect-free lattice for 1020 vortices but
observe a kagom\'e lattice with defects for 2958 vortices which may be
a consequence of the simulated annealing rate. All simulations were
initialized with random configurations and later compared with a
perfect lattice. In the case of the honeycomb and kagom\'e lattice
results, we observed a polycrystalline state which had higher energy
than the perfect lattice. To ensure that the perfect lattice was the
correct ground state, we prepared simualtions with the ground state
configuration at high temperature repeated the simulated annealing
protocol, ending up with a final configuration lower than the
defect-filled case (see Fig.~\ref{fig:snapshots}(a,d) for lowest
energy configurations).


\begin{figure}[tb]
  \centering
  \includegraphics[width=0.5\textwidth]{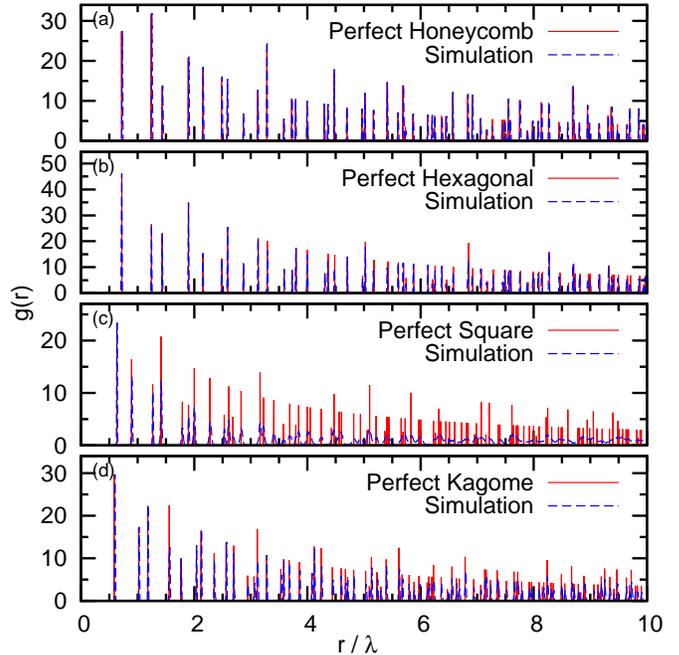}
  \caption{ 
    Comparison of the radial distribution function $g(r)$ of the vortex
    configurations shown in Fig.~\ref{fig:snapshots} with those of the
    ideal geometry for (a) honeycomb, (b) hexagonal, (c) square, and
    (d) kagom\'e lattices. The dashed blue line is the zero temperature
    result after simulated annealing, and the solid red line is the
    ideal result.
    \label{fig:rdf}
  }
\end{figure}

In order to characterize the degree of perfection for each phase, we
first consider the radial distribution function~(RDF),
\begin{align}
  g(r) = \frac{1}{2\pi r \Delta r \rho N_v} \sum_{i = 1}^{N_v} n_i(r,\Delta r),
\end{align}
where $n_i(r,\Delta r)$ is the number of particles in the shell
surrounding the $i$-th particle with radius $r$ and thickness $\Delta
r$. For phases that form regular lattice structures, we can offer a
direct comparison with an ideal lattice, which we illustrate in
Fig.~\ref{fig:rdf}.

\begin{figure}[t]
  \centering
  \includegraphics[width=0.5\textwidth]{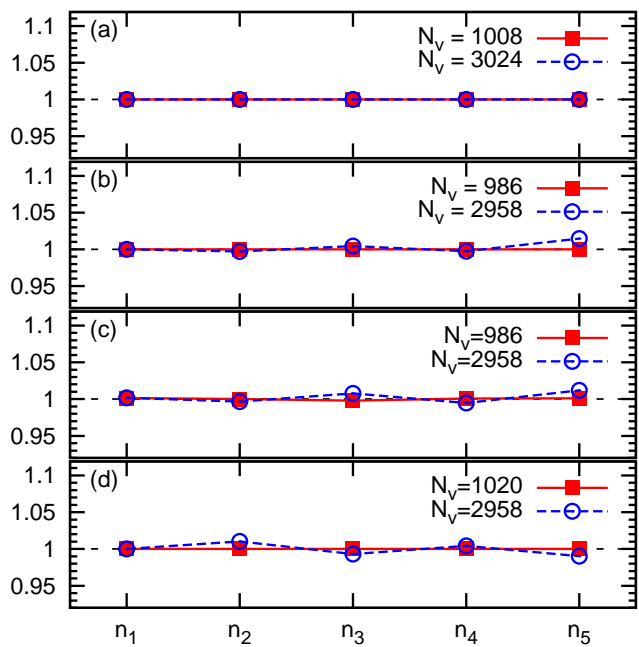}
  \caption{Number of nearest neighbors $n_i$ up to the
    fifth-nearest-neighbor for the (a) honeycomb (b) hexagonal, (c)
    square, and (d) kagom\'e lattices of Fig.~\ref{fig:snapshots} with
    $N_v \approx 1000$ (squares) and $3000$ (circles) vortices. Here, $n_i$
    is normalized to the number of neighbors in a perfect lattice.
    \label{fig:neighbors}
  }
\end{figure}

From $g(r)$ we can define the $i$-th nearest neighbor (coordination
numbers) as 
\begin{align}
  n_i = 2 \pi \rho \int_{r_{i-1}}^{r_i} g(r) dr,
\end{align}
where $r_{i-1}$ and $r_i$ are the minima surrounding the $i$th peak in
$g(r)$. In Fig.~\ref{fig:neighbors}, we show the coordination number
up to the 5th nearest neighbor for each of the lattices shown above.

Next, we define the degree of perfection $d = \frac{1}{N_v} \sum d_j$
for a lattice as
\begin{align}
  d_j &= \frac{1}{n_1} \left| \sum_{i=1}^{n_1} \left(1 - \frac{\Delta
        \theta}{\theta_{\rm perfect}} \right) \right|, & \Delta \theta &=
  |\theta_{i}-\theta_{\rm perfect}| 
\end{align}
where $d_j$ is the degree of perfection for the $j$th vortex, $n_1$ is
the number of the nearest neighbors (i.e. the number of the vortices
within a circle of radius $r_c$ with the $j$th vortex at its center, where $r_c$ is the first minimum of the RDF), $\theta_i$ is the angle
between the two nearest neighbours, and $\theta_{\rm perfect}$ is the
angle between the two nearest neighbours in the perfect lattice. Note
that by definition, $d = 1$ if there are no defects in the
lattice. For the square, hexagonal, and honeycomb lattices
$\theta_{\rm perfect} = \pi / 2$, $\pi / 3$, and $2\pi / 3$,
respectively, while the kagom\'e lattice has two possible angles: $\pi
/ 3$ and $2\pi/3$.

For the honeycomb lattice (panel (a) of Figs.~\ref{fig:snapshots},
\ref{fig:rdf}, and \ref{fig:neighbors}), we find that the ordering of
the vortices matches the ideal result very well, with the degree
of perfection $d \approx 1$ for all simulations of $N_v = 1008$ and $N_v = 3024$ vortices. The peaks of the radial distribution function closely match the ideal case, with broadening of the peaks due to defects that increases as the separation between the vortices increases. The coordination number is within 1\% for all results.

For the hexagonal lattice [panel (b)], the ordering is nearly perfect,
with $d \approx 1$ and the radial distribution function featuring
nearly delta function peaks that match with the ideal result. The
coordination number calculation also remains within 1\% of the ideal
result up to $n_5$ for simulations of $N_v = 2958$ and for all
coordination numbers we calculated for simulations of $N_v = 986$
vortices.

For the square lattice [panel (c)], the ordering is extremely good,
with $d = 0.990$ and $0.989$ for $N_v=986$ and $2958$ vortices,
respectively. The radial distribution function features delta function
peaks for the first eight peaks before broadening begins to occur. In
addition, the number of nearest neighbors calculated is within 1\% of
the ideal result for the first five neighbors.

For the kagom\'e lattice [panel (d)], the ordering is also very good,
with $d = 0.999$ and $0.946$ for $N_v = 1020$ and $2958$,
respectively. The radial distribution function of the simulation
result matches the perfect kagom\'e lattice peaks very well. The
coordination numbers are within 1\% for both $N_v=1020$ and 2958
vortices.





In summary, the recent
proposal~\cite{romero-isart_superconducting_2013} of realizing quantum
emulators by trapping ultra-cold atoms in the magnetic field of
superconducting vortex lattice raises the need to develop methods to
create vortex lattices of various symmetries. Here we propose layered
systems where vortex interaction is multi-scale (in particular the
type-1.5 systems) as the systems where in principle various vortex
lattice symmetries can be realized. The upper layer may in particular
be used to tune localization of the field while lower layers and
interlayer distances are used to control lattice symmetry. Different
temperature dependencies of components in different layers can also be
used to manipulate the vortex lattice by controlling the temperature. We
support that proposal by simulation of point-particle objects with
phenomenological two-body forces similar to long-range forces between
straight and rigid vortex lines. Next we plan to investigate it in the
layered Ginzburg-Landau model which also include the effects of vortex
bending and non-pairwise inter-vortex forces (which can be especially
important in type-1.5 regime \cite{carlstrom_type-1.5_2011}).

This work was supported by the National Science Foundation under the
CAREER Award DMR-0955902, Knut and Alice Wallenberg Foundation through
a Royal Swedish Academy of Sciences Fellowship, and by the Swedish
Research Council.

\bibliography{coldatom07}

\begin{thebibliography}{10}
\bibitem{noterho}%
  \BibitemOpen
  \bibinfo {note} {The definition of the density is: $\rho=N_v / L_x L_y$,
  where $N_v$ is the number of vortices, $L_x$ and $L_y$ are the box size of
  simulation.}%
  \bibAnnoteFile{Stop}{noterho}%
\end{thebibliography}
\end{document}